%% file: PaperSample_Guideline.tex
\documentclass[conference,a4paper]{APSIPA2026}
\usepackage{amsmath}
\usepackage{graphicx}
\usepackage{multirow}
\usepackage{threeparttable}
\usepackage[backend=biber,style=ieee,]{biblatex}
\usepackage{booktabs}
\usepackage{geometry}
\usepackage{fancyhdr}

\fancypagestyle{firststyle}{
  \fancyhf{}
}

\begin{document}

\title{RRP-Voice: A Longitudinal Dataset and Benchmark for Recurrent Respiratory Papillomatosis Detection}

\author{
  \authorblockN{
    Wenze Ren\authorrefmark{1},
    Ke-Han Lu\authorrefmark{1},
    Kai-Wei Chang\authorrefmark{4},
    Tiantian Feng\authorrefmark{7},
    Ching Fang\authorrefmark{8},
    Zhi-Chi Liao\authorrefmark{2},
    Dao Thi Hai Yen\authorrefmark{2},\\
    Syu-Siang Wang\authorrefmark{6},
    Yu Tsao\authorrefmark{3},
    Chi-Te Wang\authorrefmark{5},
    Shih-Hau Fang\authorrefmark{2}
  }
  \authorblockA{
    \authorrefmark{1}National Taiwan University,
    \authorrefmark{2}National Taiwan Normal University,
    \authorrefmark{3}Academia Sinica,
    \authorrefmark{4}Massachusetts Institute of Technology,\\
    \authorrefmark{5}Far Eastern Memorial Hospital,
    \authorrefmark{6}Yuan Ze University,
    \authorrefmark{7}University of Southern California,\\
    \authorrefmark{8}Taipei Municipal Zhongshan Girls High School
  }
}

\maketitle
\thispagestyle{firststyle}
\pagestyle{fancy}

\begin{abstract}
  % Pathological voice detection has advanced rapidly with deep learning, yet rare laryngeal diseases remain underexplored due to data scarcity. Recurrent Respiratory Papillomatosis (RRP), an HPV-induced disease marked by repeated laryngeal tumor recurrence, exemplifies this challenge: patients oscillate between active disease and post-surgical remission over years, demanding continuous voice monitoring. Yet existing pathological voice datasets are cross-sectional and fail to capture such intra-subject dynamics. We present the first longitudinal voice dataset for RRP: recordings from 26 patients with up to ten years of follow-up, comprising sustained vowels and sentence-level utterances annotated by otolaryngologists with synchronous laryngoscopic confirmation. Building on this resource, we establish a systematic benchmark spanning handcrafted features, deep networks trained from scratch, self-supervised pretrained models, and recent audio large language models, all evaluated under strict patient-independent protocols. Per-subject longitudinal analyses further verify that the cross-sectional discriminative signal reflects laryngoscopic disease state rather than stable speaker attributes. This work lays a foundation for studying rare longitudinal pathological voice tasks in low-resource clinical settings.
Deep learning has advanced pathological voice detection rapidly, yet rare laryngeal diseases remain underexplored due to data scarcity. Recurrent Respiratory Papillomatosis (RRP) exemplifies this gap: an HPV-induced disease of the larynx in which patients oscillate between recurrence and post-surgical remission over the years. RRP demands continuous voice monitoring that existing cross-sectional corpora cannot support. We introduce the first longitudinal voice dataset for RRP, comprising recordings from 26 patients with up to ten years of follow-up. Each session pairs sustained vowels with sentence-level utterances, which are annotated by otolaryngologists and confirmed synchronously with laryngoscopy. Building on this resource, we establish a systematic benchmark spanning handcrafted features, end-to-end deep networks, self-supervised pretrained models, and recent audio large language models, all evaluated under session-level cross-validation with patient-level audit. Per-subject longitudinal analyses further confirm that the cross-sectional discriminative signal reflects laryngoscopic disease state rather than stable speaker attributes. This work lays a foundation for rare longitudinal pathological voice tasks in low-resource clinical settings.

\end{abstract}

\input{Sections/1_Introduction}
\input{Sections/2_Dataset}
\input{Sections/3_Benchmark_Methods}
\input{Sections/4_Experimental_Setup}
\input{Sections/5_Cross-Sectional_Benchmark_Results}

\input{Sections/6_Longitudinal_Analysis}
\input{Sections/7_Conclusion_and_Future_Work}

% \section*{Acknowledgment}
% The preferred spelling of the word ``acknowledgment'' in America is without an ``e'' after the ``g.''
% Try to avoid the stilted expression, ``One of us (R. B. G.) thanks ...''
% Instead, try ``R.B.G. thanks ...''
% Put sponsor acknowledgments in the unnumbered footnote on the first page.

\printbibliography

\end{document}

%% file: Sections/1_Introduction.tex
\section{Introduction}

Voice encodes rich physiological information about the vocal apparatus and has emerged as a non-invasive, low-cost, and remotely accessible biomarker of laryngeal health~\cite{voiceasabiomarker, voiceforhealth}. Over the past decade, deep learning combined with public corpora such as the Saarbrücken Voice Database(SVD)~\cite{WoldertJokisz2007SaarbrueckenVD} has driven substantial progress on common laryngeal pathologies, with modeling moving from handcrafted acoustic descriptors through end-to-end deep networks to self-supervised representations and audio LLMs~\cite{TEIXEIRA20141228, spdbasedonmfcc, 8513222, harar2017voice, baevski2020wav2vec, chen2022wavlm, 10041907, team2023gemini, comanici2025gemini}.On common laryngeal pathologies such as functional voice disorders, vocal fold paralysis, and laryngeal carcinoma, these systems have already attained clinically meaningful accuracy.

% Despite this progress, two persistent gaps continue to limit the clinical reach of voice-based diagnostics. First, rare laryngeal diseases remain severely under-represented in publicly available data. Existing corpora are dominated by prevalent disorders, for which sufficient recordings can be readily assembled, whereas low-prevalence conditions, which are often the ones in greatest need of remote monitoring, rarely benefit from dedicated benchmarks. Second, virtually all existing datasets are cross-sectional in design: each patient contributes recordings at only a single time point, which are then treated as independent samples. This design implicitly assumes that pathological voice constitutes a static phenotype, and it is therefore unable to capture intra-subject dynamics, namely how voice evolves within the same individual as the disease progresses, recurs, or remits. For conditions with an intrinsically episodic clinical course, such cross-sectional snapshots discard precisely the information that matters most for longitudinal surveillance.
Despite this progress, two persistent gaps limit the clinical reach of voice-based diagnostics. First, rare laryngeal diseases remain severely under-represented in publicly available data~\cite{DepAI}. Existing corpora are dominated by prevalent disorders, for which recordings can be readily assembled, whereas low-prevalence conditions, which are often in greatest need of remote monitoring, rarely receive dedicated benchmarks~\cite{jcm14197093}. Second, virtually most existing datasets are cross-sectional: each patient contributes recordings at a single time point, treated as independent samples. This design implicitly assumes that pathological voice constitutes a static phenotype, and therefore cannot capture intra-subject dynamics~\cite{lvmdbyodt}, namely how voice evolves within an individual as the disease progresses, recurs, or remits. For conditions with an intrinsically episodic course, such snapshots discard precisely the information that matters most for longitudinal surveillance.

In this paper, we introduce the first longitudinal voice dataset dedicated to RRP. We provide a systematic benchmark and per-subject longitudinal characterization. Building on this resource, our contributions are as follows:
\begin{enumerate}
    \item \textbf{A longitudinal RRP voice corpus.} We introduce the first voice corpus dedicated to RRP, with per-subject follow-up reaching up to ten years at its longest and synchronous laryngoscopic ground truth at each visit. This resource addresses a gap left unfilled by existing pathological voice corpora.
    
    \item \textbf{A systematic benchmark across four representation families.} We evaluate handcrafted features, end-to-end deep networks, self-supervised speech models, and audio LLMs, providing a reference point for future low-resource pathological voice research.
    
    \item \textbf{Per-subject longitudinal validation.} Leveraging longitudinal observations, we verify that the cross-sectional discriminative signal tracks disease state rather than stable speaker attributes, a check infeasible on cross-sectional corpora.
\end{enumerate}

% We deliberately scope this work to cross-sectional classification baselines complemented by descriptive longitudinal analyses. Our aim is to lay the foundation for future research on RRP voice diagnostics: a clinically grounded longitudinal resource, an honest set of baselines, and an initial characterization of the within-subject voice dynamics exhibited by this disease.

% The corpus includes recordings from 26 patients at a tertiary referral center, with individual follow-ups lasting up to ten years. During these visits, we collected sustained vowels and sentence-level utterances. Each recording is annotated by board-certified otolaryngologists and paired with laryngoscopic confirmation of disease state, offering clinically grounded rather than self-reported labels. 
This work is deliberately scoped to cross-sectional classification baselines, complemented by descriptive longitudinal analyses. It aims to establish a foundation for future research on RRP voice diagnostics by providing a clinically grounded longitudinal resource, reference baselines, and an initial characterization of within-subject voice dynamics.

%% file: Sections/2_Dataset.tex
\section{Dataset}

\subsection{Clinical Setting and Recording Protocol}

% All recordings analyzed in this study are drawn from a longitudinal clinical voice corpus, assembled under a study protocol approved by the relevant institutional ethics review board. The cohort comprises 26 patients with a confirmed clinical diagnosis of Recurrent Respiratory Papillomatosis (RRP), consecutively recruited from the otolaryngology outpatient clinic. At each follow-up visit, two acoustic tasks were recorded: a sustained /a/ vowel, used to probe vocal-fold vibratory stability under steady-state phonation; and a sentence-level utterance, used to capture connected-speech characteristics under natural prosody. Because the recordings were obtained during regular clinical visits rather than in a dedicated experimental session, the dataset faithfully reflects the acoustic conditions, patient compliance, and visit cadence that any future home-monitoring system would realistically encounter.
The corpus comprises recordings from 26 patients with a confirmed clinical diagnosis of RRP, consecutively recruited from an otolaryngology outpatient clinic under a study protocol approved by the institutional ethics review board. At each follow-up visit, two acoustic tasks were administered. A sustained /a/ vowel was recorded to probe vocal-fold vibratory stability under steady-state phonation, and a sentence-level utterance was recorded to capture connected-speech characteristics under natural prosody. Because all recordings were obtained during routine clinical visits rather than in dedicated experimental sessions, the dataset preserves the acoustic conditions, patient compliance, and visit cadence that any future home-monitoring system would realistically encounter.

\subsection{Cohort Composition and Longitudinal Structure}
\input{Figures/all_longitudinal}
% The study cohort comprises 26 enrolled patients (18 males and 8 females), aged 13 to 73 years (mean 40.8). Data acquisition spanned more than a decade, from December 2013 to November 2025, yielding 151 recordings across the cohort. The number of recordings per patient varies substantially, ranging from 2 to 15, with an average of 5.8 sessions per patient and a mean follow-up duration of 3.5 years. The inter-visit interval has a mean of 266 days but a standard deviation of 405 days, indicating that the variability of these intervals exceeds their typical magnitude. Such overdispersion directly reflects the irregular, real-world scheduling that characterizes postoperative laryngology care. Figure \ref{fig:all_longitudinal} visualizes this longitudinal structure at the per-subject level, plotting the timing and clinical label of every visit across all 26 patients; the marked heterogeneity in follow-up duration, visit count, and inter-visit spacing is immediately apparent, as is the interleaving of normal and abnormal sessions within individual disease courses. Taken together, these properties define the data regime that motivates our framework. Each patient contributes only a handful of recordings, distributed unpredictably over years, against a backdrop of substantial anatomical and acoustic variation between individuals. Under such conditions, conventional supervised deep models are unlikely to generalize reliably.
The cohort consists of 18 males and 8 females, aged 13 to 73 years (mean 40.8). Data acquisition spanned more than a decade, from December 2013 to November 2025, yielding 151 recordings. The number of sessions per patient ranges from 2 to 15 (mean 5.8), with a mean follow-up duration of 3.5 years. The inter-visit interval exhibits pronounced overdispersion (mean 266 days, standard deviation 405 days), reflecting the irregular scheduling typical of postoperative laryngology care.
Figure \ref{fig:all_longitudinal} visualizes this longitudinal structure at the per-subject level, plotting the timing and clinical label of every visit across all 26 patients. The heterogeneity in follow-up duration, visit count, and inter-visit spacing is immediately apparent, as is the interleaving of normal and abnormal sessions within individual disease courses. These properties define a challenging data regime in which each patient contributes only a handful of recordings, distributed unpredictably over years, against a backdrop of substantial anatomical and acoustic variation across individuals. Under such conditions, conventional supervised deep models are unlikely to generalize reliably.

\subsection{Ground-Truth Annotation}
% Each recording is paired with a binary clinical label assigned by an otolaryngologist at the time of the visit, where 0 indicates the absence of papillomatous lesions, and 1 indicates their presence. The label is determined by the otolaryngologist's laryngoscopic examination during the same visit, so the ground truth reflects clinical observation of the larynx rather than retrospective acoustic judgment. Annotation is performed at the recording level, with a single label per visit and no within-recording temporal segmentation. Across the 151 recordings in the corpus, the two classes are distributed at approximately 57\% (label 0) and 43\% (label 1).
Each recording is paired with a binary clinical label assigned by an otolaryngologist at the time of the visit: 0 denotes the absence of papillomatous lesions, and 1 denotes their presence. Labels are determined by laryngoscopic examination performed during the same visit, so the ground truth reflects direct clinical observation of the larynx rather than retrospective acoustic judgment. Annotation is conducted at the recording level, with a single label per visit and no within-recording temporal segmentation. Across the 151 recordings, the two classes are distributed at approximately 57\% (label 0) and 43\% (label 1).

%% file: Figures/all_longitudinal.tex
\begin{figure}[t]
\centering
\caption{Per-subject longitudinal visit timeline for the 26 patients, where each vertical bar marks a clinical visit colored by laryngoscopic label.}
% \vspace{-6pt}
\includegraphics[width=\columnwidth]{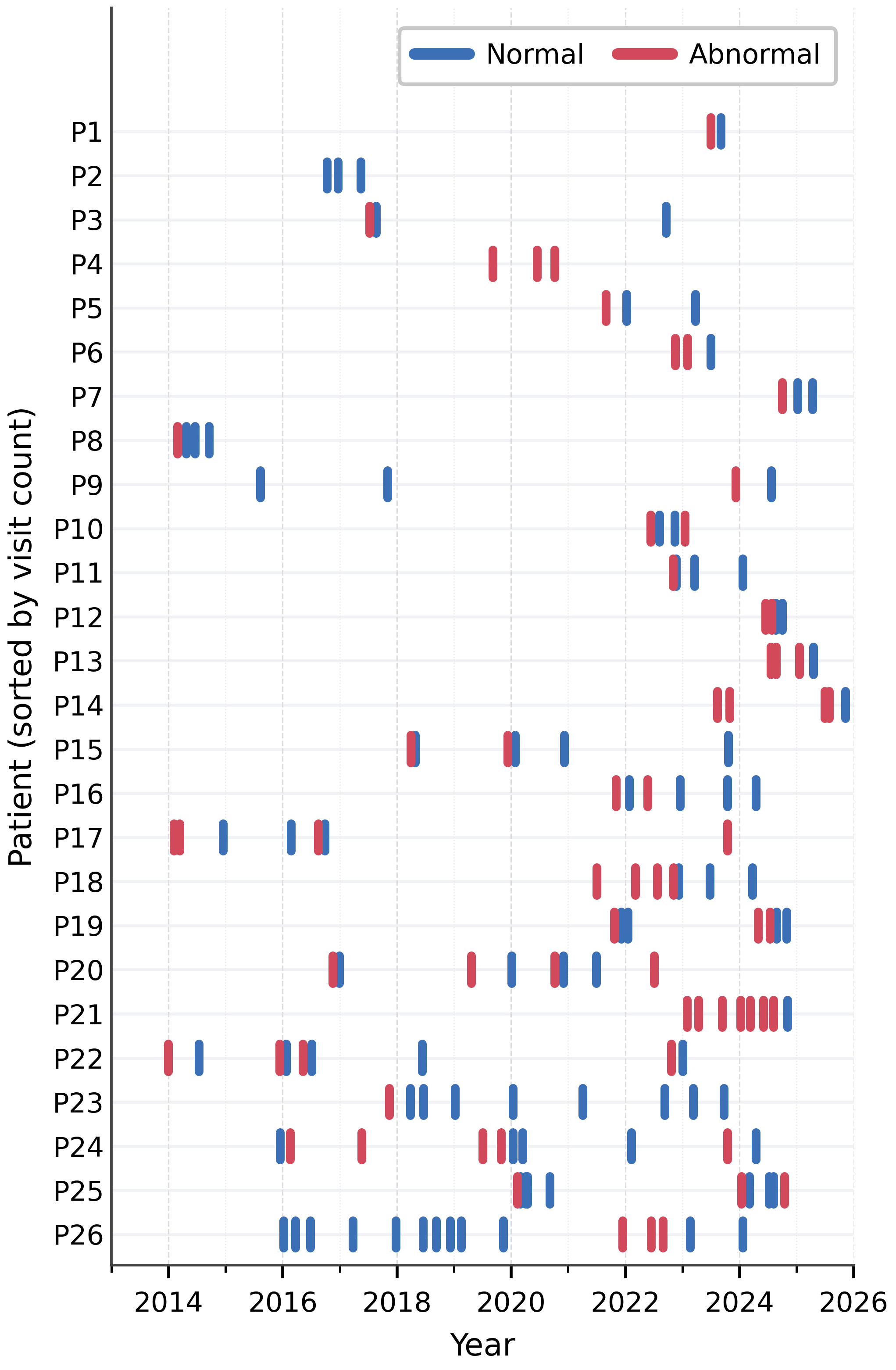}
\vspace{-32pt}
\label{fig:all_longitudinal}
\end{figure}

%% file: Sections/3_Benchmark_Methods.tex
\section{Benchmark Methods}
\label{sec:methods}

We benchmark four representation families spanning the methodological history of pathological voice detection: (i) handcrafted acoustic functionals with a gradient-boosted tree; (ii) a compact CNN trained end-to-end on log-Mel spectrograms; (iii) frozen self-supervised waveform embeddings with a lightweight MLP; and (iv) audio large language models (LLMs). Since each visit yields a paired sustained vowel and sentence, the first three families are evaluated in vowel-only, sentence-only, and fusion configurations; the fourth is evaluated only by recording type.

\subsection{eGeMAPS + LightGBM}
Each recording is encoded by the 88-dimensional eGeMAPSv02~\cite{7160715} functional set and classified by LightGBM~\cite{NIPS2017_6449f44a}(200 trees, 15 leaves, lr=0.05, min five samples per leaf, class-balanced weights). The fusion variant concatenates the session's vowel and sentence vectors into a 176-d input.

\subsection{Log-Mel CNN trained from scratch}
We refer to this baseline as \emph{SmallMel-CNN}: a compact CNN tower operating on a 128-bin log-Mel patch extracted from a three-second waveform crop. It stacks four Conv-BN-ReLU blocks (channels $\{16,32,64,128\}$, $3{\times}3$ kernels), with $2{\times}2$ max-pooling on the first three blocks and adaptive average pooling on the last; a dropout layer ($p{=}0.3$) and a linear classifier map the resulting 128-d embedding to a logit. The fusion variant instantiates two independent SmallMel-CNN towers, one per recording type, whose 128-d embeddings are concatenated and passed to a two-layer MLP ($256{\to}64{\to}1$, with dropout).

\subsection{Frozen wav2vec\,2.0 features + MLP}
Each three-second waveform is passed through a frozen wav2vec 2.0 base backbone and mean-pooled along time, yielding a 768-d embedding that a two-layer MLP ($768{\to}256{\to}1$, ReLU, dropout 0.2) maps to a logit. For fusion, the same backbone encodes both streams of a session; their embeddings are concatenated and fed to an MLP of identical depth ($1536{\to}256{\to}1$).

\subsection{Audio large language models}
We query two audio-capable Gemini LLMs, Gemini 2.5 Flash and Gemini 3.1 Pro Preview, with the raw WAV and a fixed two-part prompt. The system prompt frames the model as an Ear, Nose, and Throat (ENT) specialist performing binary RRP screening; the user prompt provides the recording type, the patient's self-reported age and gender, and clinically relevant acoustic cues (hoarseness, roughness, and breathiness for vowels; vocal break and strain for sentences). The model returns a JSON object with a binary label, a confidence in $[0,1]$, and a brief explanation. Vowel and sentence recordings are classified independently; no fusion variant is considered.

%% file: Sections/4_Experimental_Setup.tex
\section{Experimental Setup}

\subsection{Cross-validation and audio processing}
All models are evaluated under a single 5-fold cross-validation split 
generated with a fixed seed, ensuring direct comparability across method families. The atomic unit of partitioning is the \emph{session}: paired vowel and sentence recordings from the same visit are jointly assigned to one fold, preserving session integrity for fusion variants. All recordings are resampled to 16~kHz mono. Subject-level leave-one-out is infeasible given the cohort size and the presence of single-class patient sequences; the resulting speaker-identity confound is audited empirically in Section~\ref{sec:longitudinal}. The eGeMAPS baseline and the audio LLMs consume the entire utterance, whereas the two trained-from-scratch baselines (CNN and wav2vec\,2.0) operate on fixed three-second windows: a single random crop per file per epoch during training (zero-padded if shorter) and non-overlapping segments at inference. File-level probabilities for single-stream variants are obtained by averaging per-segment sigmoid scores, while fusion variants first average segment embeddings within each stream and then concatenate the two stream-level vectors for a single forward pass through the head. For the log-Mel CNN, 128 Mel bands are extracted with a 25~ms Hann window and a 10~ms hop ($n_{\text{FFT}}{=}512$), followed by a natural log of the power spectrogram.

\subsection{Training and decision rules}
The two neural baselines follow an identical optimization recipe: AdamW 
($\text{lr}{=}10^{-3}$, weight decay $10^{-4}$, $\beta{=}(0.9, 0.999)$) 
for 25 epochs, with a two-epoch linear warm-up followed by cosine annealing, gradients clipped at norm $1.0$, and binary cross-entropy loss. The batch size is $32$ for single-stream variants and $16$ for fusion, and all training is performed on a single NVIDIA RTX~4090 GPU. The eGeMAPS$+$LightGBM model is fit once per fold, whereas the audio LLMs are queried zero-shot, with their predictions aggregated within each fold to match the per-fold reporting of supervised baselines. Two decision rules are reported for the supervised methods: a fixed threshold of $0.5$, and a per-fold threshold selected by grid search over $\{0.05, 0.06, \ldots, 0.95\}$ that maximizes UAR on training-partition predictions and is then applied unchanged to the held-out split, so that test labels are never accessed during threshold selection. The audio LLMs output discrete labels and are reported under their native decisions.

\subsection{Evaluation metrics}
Given the mild class imbalance (57\% normal, 43\% abnormal) and the screening nature of the task, our primary metric is \textbf{unweighted average recall}(UAR), defined as the unweighted mean of sensitivity and specificity. We additionally report sensitivity, specificity, the $F_1$ score on the abnormal class, accuracy, and the threshold-free AUC-ROC. Within each fold, metrics are computed at the file level for single-stream variants and at the session level for fusion, and final results are reported as mean$\pm$std across the five folds.

%% file: Sections/5_Cross-Sectional_Benchmark_Results.tex
\section{Cross-Sectional Benchmark Results}
\input{Tables/All_model_results}
% Table~\ref{tab:all-results} reports the unified five-fold
% cross-validation results across the four representation families.
% Three observations frame our interpretation of the benchmark.
Table~\ref{tab:all-results} reports the unified five-fold cross-validation results across the four representation families. Three observations frame our interpretation.

\subsection{Self-supervised pretraining dominates the supervised baselines}
% Across all input configurations and decision rules, frozen wav2vec~2.0 features paired with a lightweight MLP achieve the highest UAR and AUC-ROC among the supervised baselines. Under the fixed threshold $\tau=0.5$, the fusion variant reaches UAR $0.787 \pm 0.028$ and AUC-ROC $0.866 \pm 0.026$, exceeding the best eGeMAPS+LightGBM configuration (UAR $0.739$, AUC-ROC $0.811$) by $4.8$ UAR points and the log-Mel CNN trained from scratch (UAR $0.656$, AUC-ROC $0.712$) by $13.1$ UAR points. The CNN remains the weakest baseline across nearly every metric in Panel~(a), consistent with the cohort scale: with only $151$ recordings from $26$ patients, an end-to-end network has too few examples to learn discriminative spectrotemporal filters under the strict patient-independent protocol. The self-supervised backbone, by contrast, supplies representations that already encode broad phonatory regularities from large-scale unlabeled speech, leaving only a low-capacity head to be fit on in-domain data. This pattern echoes prior findings on common laryngeal pathologies and suggests that, for rare-disease voice tasks where data scarcity is structural rather than incidental, transfer from self-supervised speech models should be the default rather than an optional refinement.
Frozen wav2vec~2.0 features paired with a lightweight MLP achieve the highest UAR and AUC-ROC among the supervised baselines across all input configurations and decision rules. Under the fixed threshold $\tau=0.5$, the fusion variant reaches UAR $0.787 \pm 0.028$ and AUC-ROC $0.866 \pm 0.026$, exceeding the best eGeMAPS+LightGBM configuration (UAR $0.739$, AUC-ROC $0.811$) by 4.8 UAR points and the log-Mel CNN (UAR $0.656$, AUC-ROC $0.712$) by 13.1 UAR points. The CNN remains the weakest baseline in Panel~(a), a result consistent with the cohort scale: with only 151 recordings from 26 patients, an end-to-end network has too few examples to learn discriminative spectrotemporal filters. In contrast, the self-supervised backbone already encodes broad phonatory regularities from large-scale unlabeled speech, leaving only a low-capacity head to be fit on in-domain data. This pattern echoes prior findings on common laryngeal pathologies. For rare-disease voice tasks in which data scarcity is structural rather than incidental, transfer from self-supervised speech models should therefore be the default rather than an optional refinement. 

\subsection{Fusion gains scale with representation quality}
% Fusing the paired vowel and sentence recordings improves UAR over the better single-stream variant for the two stronger representation families. The gain is most pronounced for wav2vec~2.0+MLP: fusion lifts UAR from $0.726$ (sentence, the better single stream) to $0.787$, a $6.1$-point absolute improvement, and AUC-ROC from $0.814$ to $0.866$. eGeMAPS shows a smaller but consistent gain ($0.715 \rightarrow 0.739$). The log-Mel CNN, by contrast, fails to benefit reliably: its AUC-ROC slips from $0.721$ (vowel) to $0.712$ (fusion) under the fixed threshold, and once the threshold is tuned, UAR itself degrades from $0.657$ to $0.630$.

% This asymmetry is informative. The two phonatory contexts capture genuinely complementary cues: sustained vowels probe the steady-state vibratory stability of the vocal folds, whereas sentences expose prosodic vocal breaks and strain under connected speech dynamics. Exploiting these cues jointly, however, requires a representation expressive enough to encode them; weak representations have little to fuse, while strong ones compound their gains. The fusion result, therefore, validates the clinic's dual-task recording protocol and identifies multi-task acoustic elicitation as a low-cost lever for improving rare-disease voice detection.
Fusing the paired vowel and sentence recordings improves UAR over the better single-stream variant for the two stronger representation families. The gain is most pronounced for wav2vec~2.0+MLP, where fusion lifts UAR from $0.726$ (sentence) to $0.787$, a 6.1-point absolute improvement, and AUC-ROC from $0.814$ to $0.866$. eGeMAPS shows a smaller but consistent gain ($0.715$ to $0.739$). The log-Mel CNN, by contrast, fails to benefit reliably: its AUC-ROC drops from $0.721$ (vowel) to $0.712$ (fusion) at the fixed threshold, and further threshold tuning degrades UAR from $0.657$ to $0.630$. 
% \vspace{-pt}
This asymmetry is informative. Sustained vowels probe the steady-state vibratory stability of the vocal folds, whereas sentences expose prosodic vocal breaks and strain under connected-speech dynamics. Exploiting these complementary cues jointly requires a representation expressive enough to encode them: weak representations have little to fuse, whereas strong ones compound their gains. The fusion result therefore, validates the clinic's dual-task recording protocol and identifies multi-task acoustic elicitation as a low-cost lever for rare-disease voice detection.
\vspace{-1.5pt}
\subsection{Zero-shot audio LLMs underperform and exhibit failure modes}
% Despite their broad capabilities, the audio LLMs evaluated zero-shot underperform every supervised baseline. The strongest configuration, Gemini~3.1~Pro~Preview on sentences, attains UAR $0.652 \pm 0.090$, trailing the simplest supervised fusion baseline (eGeMAPS+LightGBM, UAR $0.739$) by $8.7$ points and the best supervised model (wav2vec~2.0 fusion, UAR $0.787$) by $13.5$ points. The shortfall is also qualitative: Gemini~2.5~Flash collapses on sustained vowels into a degenerate predictor that labels every recording as abnormal (sensitivity $1.0$, specificity $0.0$), a failure mode no supervised baseline reproduces. Even when the LLMs avoid such degeneracy, their predictions on vowels remain markedly less reliable than on sentences, suggesting that the narrative-text priors on which these models are trained generalize poorly to short, lexically empty acoustic stimuli.

% Taken together, these results indicate that current audio LLMs cannot yet substitute for task-specific adaptation in rare pathological voice detection: the acoustic signatures of RRP are subtle, intra-class variability is large, and laryngoscopically grounded supervision remains essential. They nonetheless establish a meaningful zero-shot reference and motivate future work on instruction tuning or in-context demonstration strategies tailored to longitudinal voice diagnostics.
Despite their broad capabilities, the zero-shot audio LLMs underperform every supervised baseline. The strongest configuration, Gemini~3.1~Pro~Preview on sentences, attains UAR $0.652 \pm 0.090$, trailing the eGeMAPS+LightGBM fusion baseline (UAR $0.739$) by $8.7$ points and the wav2vec~2.0 fusion model (UAR $0.787$) by 13.5 points. The shortfall is also qualitative. Gemini~2.5~Flash collapses on sustained vowels into a degenerate predictor that labels every recording as abnormal (sensitivity $1.0$, specificity $0.0$), a failure mode no supervised baseline reproduces. Even when the LLMs avoid such degeneracy, their predictions on vowels remain markedly less reliable than on sentences, suggesting that the narrative-text priors on which these models are trained generalize poorly to short, lexically empty acoustic stimuli.
\vspace{-1pt}
Taken together, these results indicate that current audio LLMs cannot yet substitute for task-specific adaptation in rare pathological voice detection. The acoustic signatures of RRP are subtle, intra-class variability is large, and laryngoscopically grounded supervision remains essential. The zero-shot results nonetheless establish a meaningful reference point and motivate future work on instruction tuning or in-context demonstration strategies tailored to longitudinal voice diagnostics.

%% file: Tables/All_model_results.tex
% =============================================================================
% Required preamble:
%   \usepackage{booktabs}   % \toprule, \midrule, \cmidrule, \bottomrule
%   \usepackage{multirow}   % \multirow for method-name grouping
% =============================================================================

\begin{table*}[t]
\centering
\caption{Unified five-fold cross-validation benchmark on the RRP-Voice corpus.
Panels~(a) and~(b) report the three supervised baselines under the two decision rules,
while panel~(c) reports the zero-shot audio LLMs, which output discrete labels and therefore admit neither AUC-ROC (denoted ``\texttt{-}'') nor threshold tuning.
Each cell shows mean\,$\pm$\,std across the five folds, with the column-wise maximum within each panel highlighted in \textbf{bold}.
Because AUC-ROC is threshold-free, its values in panel~(b) are identical to those in panel~(a) for the same row.}
\label{tab:all-results}
\setlength{\tabcolsep}{4.5pt}
\renewcommand{\arraystretch}{1.08}
\begin{tabular}{l l c c c c c c}
\toprule
\textbf{Method} & \textbf{Variant} &
\textbf{UAR} & \textbf{Sensitivity} & \textbf{Specificity} & \textbf{F\textsubscript{1}} &
\textbf{AUC-ROC} & \textbf{Accuracy} \\
\midrule
\multicolumn{8}{@{}l}{\textit{(a) Supervised baselines --- fixed threshold $\tau = 0.5$}} \\
\cmidrule(lr){1-8}
\multirow{3}{*}{eGeMAPS + LightGBM}
  & Vowel    & $0.715 \pm 0.043$ & $0.662 \pm 0.134$ & $0.769 \pm 0.117$ & $0.667 \pm 0.065$ & $0.750 \pm 0.060$ & $0.722 \pm 0.043$ \\
  & Sentence & $0.678 \pm 0.062$ & $0.554 \pm 0.058$ & $0.803 \pm 0.121$ & $0.613 \pm 0.058$ & $0.782 \pm 0.108$ & $0.695 \pm 0.068$ \\
  & Fusion   & $0.739 \pm 0.085$ & $0.662 \pm 0.115$ & $\mathbf{0.816 \pm 0.132}$ & $0.694 \pm 0.103$ & $0.811 \pm 0.067$ & $0.749 \pm 0.088$ \\
\cmidrule(lr){1-8}
\multirow{3}{*}{Log-Mel CNN}
  & Vowel    & $0.628 \pm 0.048$ & $0.615 \pm 0.129$ & $0.640 \pm 0.040$ & $0.582 \pm 0.081$ & $0.721 \pm 0.030$ & $0.629 \pm 0.037$ \\
  & Sentence & $0.586 \pm 0.094$ & $0.369 \pm 0.255$ & $0.802 \pm 0.089$ & $0.402 \pm 0.240$ & $0.680 \pm 0.091$ & $0.616 \pm 0.073$ \\
  & Fusion   & $0.656 \pm 0.036$ & $0.615 \pm 0.129$ & $0.697 \pm 0.070$ & $0.603 \pm 0.071$ & $0.712 \pm 0.046$ & $0.662 \pm 0.026$ \\
\cmidrule(lr){1-8}
\multirow{3}{*}{wav2vec\,2.0 (frozen) + MLP}
  & Vowel    & $0.711 \pm 0.044$ & $0.677 \pm 0.090$ & $0.745 \pm 0.106$ & $0.672 \pm 0.053$ & $0.803 \pm 0.050$ & $0.716 \pm 0.048$ \\
  & Sentence & $0.726 \pm 0.078$ & $0.662 \pm 0.125$ & $0.791 \pm 0.088$ & $0.680 \pm 0.097$ & $0.814 \pm 0.058$ & $0.736 \pm 0.074$ \\
  & Fusion   & $\mathbf{0.787 \pm 0.028}$ & $\mathbf{0.769 \pm 0.097}$ & $0.804 \pm 0.097$ & $\mathbf{0.757 \pm 0.036}$ & $\mathbf{0.866 \pm 0.026}$ & $\mathbf{0.788 \pm 0.031}$ \\
\midrule
\multicolumn{8}{@{}l}{\textit{(b) Supervised baselines --- threshold tuned on training partition (per fold)}} \\
\cmidrule(lr){1-8}
\multirow{3}{*}{eGeMAPS + LightGBM}
  & Vowel    & $0.678 \pm 0.078$ & $\mathbf{0.785 \pm 0.132}$ & $0.571 \pm 0.090$ & $0.664 \pm 0.085$ & $0.750 \pm 0.060$ & $0.662 \pm 0.075$ \\
  & Sentence & $0.711 \pm 0.105$ & $0.769 \pm 0.084$ & $0.652 \pm 0.150$ & $0.694 \pm 0.097$ & $0.782 \pm 0.108$ & $0.702 \pm 0.110$ \\
  & Fusion   & $0.731 \pm 0.115$ & $\mathbf{0.785 \pm 0.102}$ & $0.677 \pm 0.186$ & $0.715 \pm 0.105$ & $0.811 \pm 0.067$ & $0.722 \pm 0.123$ \\
\cmidrule(lr){1-8}
\multirow{3}{*}{Log-Mel CNN}
  & Vowel    & $0.652 \pm 0.046$ & $0.631 \pm 0.141$ & $0.673 \pm 0.129$ & $0.605 \pm 0.073$ & $0.721 \pm 0.030$ & $0.655 \pm 0.048$ \\
  & Sentence & $0.657 \pm 0.078$ & $0.662 \pm 0.078$ & $0.652 \pm 0.119$ & $0.625 \pm 0.078$ & $0.680 \pm 0.091$ & $0.656 \pm 0.082$ \\
  & Fusion   & $0.630 \pm 0.040$ & $0.539 \pm 0.161$ & $0.721 \pm 0.094$ & $0.551 \pm 0.096$ & $0.712 \pm 0.046$ & $0.642 \pm 0.028$ \\
\cmidrule(lr){1-8}
\multirow{3}{*}{wav2vec\,2.0 (frozen) + MLP}
  & Vowel    & $0.717 \pm 0.038$ & $0.723 \pm 0.062$ & $0.711 \pm 0.119$ & $0.688 \pm 0.031$ & $0.803 \pm 0.050$ & $0.716 \pm 0.048$ \\
  & Sentence & $0.696 \pm 0.094$ & $0.692 \pm 0.146$ & $0.699 \pm 0.110$ & $0.659 \pm 0.108$ & $0.814 \pm 0.058$ & $0.696 \pm 0.092$ \\
  & Fusion   & $\mathbf{0.750 \pm 0.050}$ & $0.754 \pm 0.123$ & $\mathbf{0.746 \pm 0.154}$ & $\mathbf{0.720 \pm 0.057}$ & $\mathbf{0.866 \pm 0.026}$ & $\mathbf{0.749 \pm 0.059}$ \\
\midrule
\multicolumn{8}{@{}l}{\textit{(c) Zero-shot audio LLMs --- no training, no threshold tuning}} \\
\cmidrule(lr){1-8}
\multirow{2}{*}{Gemini~2.5~Flash}
  & Sentence    & $0.565 \pm 0.083$ & $0.862 \pm 0.084$ & $0.269 \pm 0.100$ & $0.610 \pm 0.066$ & --- & $0.524 \pm 0.086$ \\
  & Vowel & $0.500 \pm 0.000$ & $\mathbf{1.000 \pm 0.000}$\,$^{\dagger}$ & $\mathbf{0.000 \pm 0.000}$\,$^{\dagger}$ & $0.602 \pm 0.006$ & --- & $0.431 \pm 0.006$ \\
\cmidrule(lr){1-8}
\multirow{2}{*}{Gemini~3.1~Pro Preview}
  & Vowel    & $0.543 \pm 0.129$ & $0.646 \pm 0.185$ & $0.441 \pm 0.097$ & $0.538 \pm 0.132$ & --- & $0.529 \pm 0.122$ \\
  & Sentence & $\mathbf{0.652 \pm 0.090}$ & $0.723 \pm 0.140$ & $\mathbf{0.582 \pm 0.094}$ & $\mathbf{0.633 \pm 0.097}$ & --- & $\mathbf{0.642 \pm 0.086}$ \\
\bottomrule
\addlinespace[2pt]
\multicolumn{8}{@{}l}{\footnotesize $^{\dagger}$\,Degenerate output: Gemini~2.5~Flash predicts the positive class for every vowel input, yielding Sensitivity$=1$ and Specificity$=0$.} \\
\end{tabular}
\end{table*}

%% file: Sections/6_Longitudinal_Analysis.tex
\section{Per-Subject Longitudinal Validation}
\label{sec:longitudinal}
\input{Figures/within_subject_auc}
% A persistent but rarely audited concern in cross-sectional pathological voice modeling is whether a high patient-independent AUC captures an intrinsic acoustic signature of disease, or merely a correlate of speaker attributes that covary with disease prevalence in the training cohort. Because each patient contributes only one sample to the test partition, the two hypotheses yield observationally indistinguishable confusion matrices: a model that learns ``a hoarse middle-aged voice is usually pathological'' scores identically to one that learns the laryngeal pathology itself. The multi-visit structure of our corpus enables a direct audit of this confound.
A persistent but rarely audited concern in cross-sectional pathological voice modeling is whether a high patient-independent AUC reflects an intrinsic acoustic signature of disease or merely a correlate of speaker attributes that covary with disease prevalence in the cohort. Because each patient contributes only one sample to the test partition, the two hypotheses yield observationally indistinguishable confusion matrices: a model that learns ''a hoarse middle-aged voice is usually pathological'' scores identically to one that learns the laryngeal pathology itself. The multi-visit structure of our corpus enables a direct audit of this confound.
\subsection{Within-Subject AUC}
For a patient $p$ with $n_+ \geq 1$ pathological and $n_- \geq 1$ normal visits, the within-subject AUC is defined as
\begin{equation}
\mathrm{AUC}_{\mathrm{ws}}(p) = \frac{1}{n_+ n_-} \sum_{i\,:\,y_i = 1} \sum_{j\,:\,y_j = 0} \mathbf{1}\!\left[\hat{s}_i > \hat{s}_j\right],
\label{eq:wsauc}
\end{equation}
where $\hat{s}_v$ is the fold-out probability assigned to visit $v$ by the wav2vec~2.0 fusion model, with both indices restricted to visits of patient $p$. Because each (pathological, normal) pair is contributed by a single subject, anatomical, demographic, and habitual phonatory attributes are held constant by construction; the only systematic source of variation within a pair is laryngoscopic state. Being undefined on cross-sectional corpora, this metric is a property unique to longitudinal data.

\subsection{Results}
Figure~\ref{fig:within_subject_auc} summarises the audit. Panel~(A) illustrates the procedure on subject~S04, whose seven visits across a decade span four laryngoscopically confirmed recurrences and three remissions. The wav2vec~2.0 fusion model correctly orders 11 of the 12 possible pairs, yielding $\mathrm{AUC}_{\mathrm{ws}}(\text{S04})=0.917$. Panel~(B) extends the analysis to the full cohort. Of the 26 patients, 24 qualify (the remaining two contribute single-class sequences), with median $\mathrm{AUC}_{\mathrm{ws}} = 1.000$. Among these qualifying patients, 21/24 ($88\%$) exceed chance and $17/24$ ($71\%$) match or exceed the patient-independent AUC of $0.847$. Marker area encodes the limited pairwise-comparison count that drives the concentration at $\mathrm{AUC}_{\mathrm{ws}}=1$.

Two conclusions follow. First, the speaker-identity confound is not a substantive driver of the benchmark. With speaker characteristics held constant within a patient, the model preserves its discriminative ranking in $88\%$ of cases. The patient-independent AUC in Table~\ref{fig:all_longitudinal} should therefore be read as a \emph{lower bound} on the model's intrinsic state-discrimination capacity, with inter-subject variation contributing noise that within-subject evaluation removes. Second, three patients (S10, S22, S24) fall at or below chance, demarcating a subpopulation for whom voice alone is insufficient as a biomarker and motivating personalized, temporally aware monitoring as future work.

We deliberately scope the present analysis to per-patient validation rather than longitudinal predictive modeling. The cohort's median of five sessions per patient and overdispersed inter-visit intervals do not yet warrant sequence-level supervised learning. Establishing that the cross-sectional signal generalizes within-subject is nonetheless a necessary precondition for any such future system.

%% file: Figures/within_subject_auc.tex
\begin{figure}[t]
\centering
\caption{Per-subject longitudinal validation. \textbf{(A)} Subject 
S04 case: ground truth (top, red: pathological; blue: normal) and 
fold-out $\hat{s}_v$ per visit (bottom); $11/12$ $P$--$N$ pairs 
correctly ordered, $\mathrm{AUC}_{\mathrm{ws}}=0.917$. 
\textbf{(B)} Within-subject AUC across the $24/26$ qualifying 
patients, sorted descending; marker area scales with the number of 
$P$--$N$ comparisons. Vertical lines: chance ($0.5$, dashed), 
patient-independent AUC ($0.847$, dash-dot), cohort median 
($1.000$, solid). S16 and S20 are excluded (single-class).}
% \vspace{-6pt}
\includegraphics[width=\columnwidth]{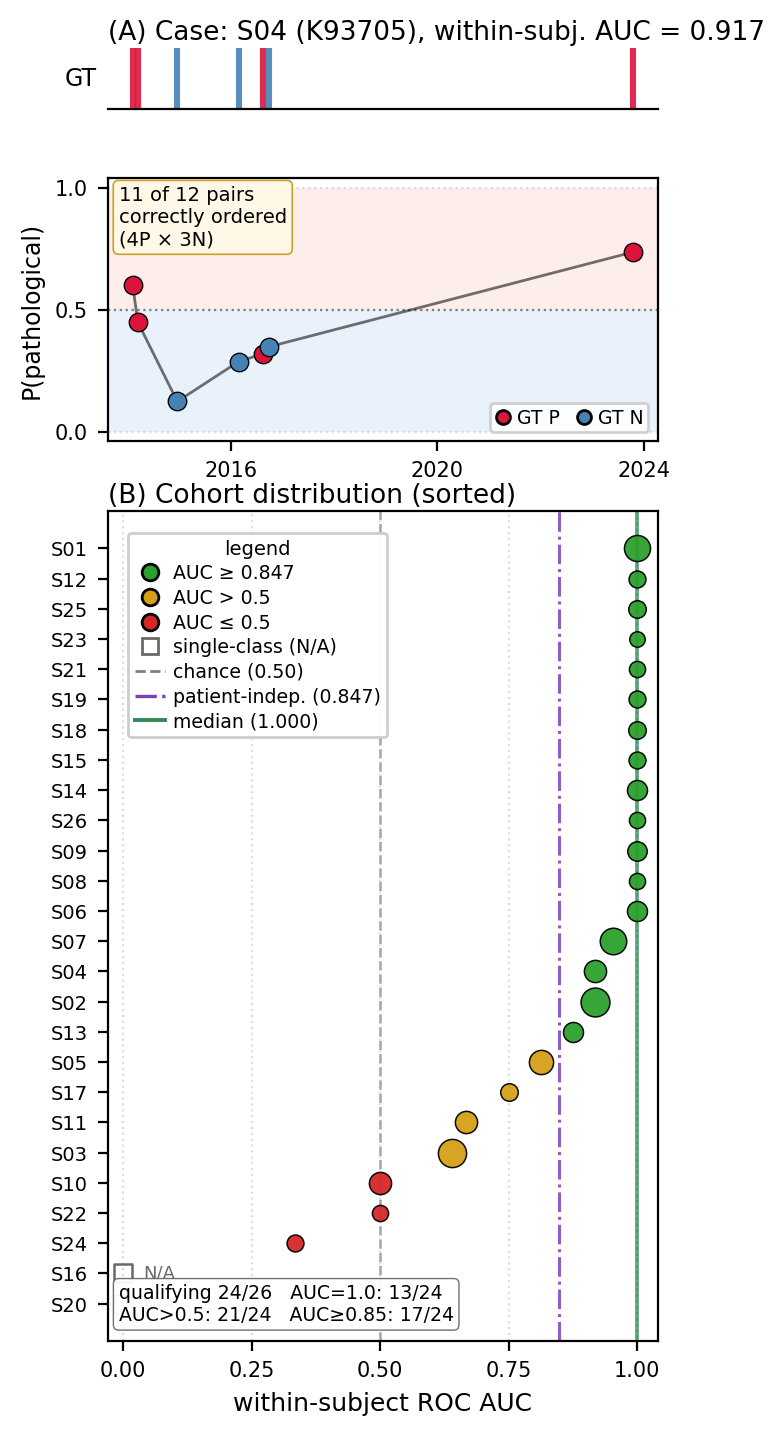}
% \vspace{-32pt}
\label{fig:within_subject_auc}
\end{figure}

%% file: Sections/7_Conclusion_and_Future_Work.tex
\section{Conclusion and Future work}
% We introduced RRP-Voice, the first longitudinal voice corpus dedicated to Recurrent Respiratory Papillomatosis, pairing every visit with synchronous laryngoscopic ground truth across up to a decade of follow-up. Benchmarking four representation families that span the methodological history of pathological voice detection (handcrafted functionals, end-to-end CNNs, self-supervised speech models, and audio LLMs) yields three findings. First, self-supervised pretraining is the most reliable inductive bias when data scarcity is structural rather than incidental. Second, zero-shot audio LLMs are not yet competitive with task-specific baselines and exhibit non-trivial failure modes on short, lexically empty stimuli. Third, a per-subject longitudinal audit confirms that the cross-sectional discriminative signal reflects disease state rather than stable speaker attributes, a verification cross-sectional corpora cannot support by construction.

% Several directions remain open: denser, more regular sampling to enable sequence-level temporal modelling and pre-clinical early warning; personalised monitoring adapted to per-subject acoustic baselines for the subpopulation whose voice fails to track laryngoscopic state; and instruction tuning or in-context demonstration to close the gap between zero-shot audio LLMs and supervised baselines. We release RRP-Voice as a clinically grounded foundation for these directions and, more broadly, for the study of episodic laryngeal disease through voice.

We introduced \emph{RRP-Voice}, the first longitudinal voice corpus for RRP, with synchronous laryngoscopic labels across up to a decade of follow-up. Benchmarking four representation families yields three findings: (i) self-supervised pretraining is the most reliable inductive bias under structural data scarcity; (ii) audio LLMs remain uncompetitive with task-specific baselines and fail on short, lexically empty stimuli; and (iii) a per-subject longitudinal audit confirms that the cross-sectional signal tracks disease state rather than speaker identity, a verification infeasible on cross-sectional corpora.

Open directions include denser sampling for sequence-level temporal modeling, personalized monitoring for subjects whose voice decouples from laryngoscopic state, and instruction tuning to close the LLM gap. We release RRP-Voice as a foundation for studying episodic laryngeal disease through voice.